\documentclass[reprint, superscriptaddress, prb, floatfix]{revtex4-2}
\usepackage{graphicx} 

\begin{document}

\title{Elastic energy driven multivariant selection in martensites via quantum annealing}

\author{Lara C. P. dos Santos}
\affiliation{Structure and Function of Materials, Institute of Energy and Climate Research IEK-2, Forschungszentrum J\"ulich GmbH, 52425 J\"ulich, Germany}

\author{Tian Hang}
\affiliation{Structure and Function of Materials, Institute of Energy and Climate Research IEK-2, Forschungszentrum J\"ulich GmbH, 52425 J\"ulich, Germany}

\author{Roland Sandt}
\affiliation{Structure and Function of Materials, Institute of Energy and Climate Research IEK-2, Forschungszentrum J\"ulich GmbH, 52425 J\"ulich, Germany}

\author{Martin Finsterbusch}
\affiliation{Materials Synthesis and Processing, Institute of Energy and Climate Research IEK-1, Forschungszentrum J\"ulich GmbH, 52425 J\"ulich, Germany}

\author{Yann Le Bouar}
\affiliation{Universit\'e Paris-Saclay, ONERA, CNRS, Laboratoire d’Etude des Microstructures, 92320 Châtillon, France}

\author{Robert Spatschek}
\affiliation{Structure and Function of Materials, Institute of Energy and Climate Research IEK-2, Forschungszentrum J\"ulich GmbH, 52425 J\"ulich, Germany}
\affiliation{JARA-ENERGY, 52425 J\"ulich, Germany}

\date{\today}

\begin{abstract}
We demonstrate the use of quantum annealing for the selection of multiple martensite variants in a microstructure with long-range coherency stresses and external mechanical load.
The general approach is illustrated for martensites with four different variants, based on the minimization of the linear elastic energy.
The equilibrium variant distribution is then analysed under application of tensile and shear strains and for different values of the considered shear and tetragonal contributions of the different martensite variants.
The interface orientations between different domains of variants can be explained using the perspective of the elastic energy anisotropy for regular stripe patterns.
For random grain orientations, the response to an external elastic strain is weaker and variants changes can be interpreted based on the rotated eigenstrain tensor.
\end{abstract}

\maketitle

\section{Introduction}

The formation of microstructures is critical for many applications, as it strongly impacts the properties of materials.
In many cases, simulation approaches can nowadays be used to support experimental investigations related to the understanding and prediction of  microstructures.
Among these approaches the phase field approach plays a central role, as it is able to predict the temporal evolution of non-equilibrium microstructures \cite{Karma1, Karma2, Boussinot2014, KaiWang2020, Finel2018}.
To this end, continuous order parameter fields are used to discriminate between different phases, grains and in particular martensite variants, and in this way phase transformation kinetics can be simulated.
Contrary to this approach, we recently presented an approach which complements the phase field picture in the sense that it separates discrete from continuous degrees of freedom \cite{Sandt2023a}.
Specific applications are stress or strain driven transformations in shape memory alloys { or martensitic transformations in general}.
There, we assumed for simplicity that a grain can be in one out of two martensite variant states, which differ by their eigenstrain, i.e.~their stress free configuration.
By application of external stresses and the presence of internal stresses, the grains can flip between the two variants, and these transitions are driven by a reduction of the overall elastic energy.

{The prediction of martensite formation and variant selection has been widely discussed in the literature, see e.g.~\cite{Mao2019}.
In this context, the concept of a lattice correspondence between austenite and martensite gained attention, resulting e.g.~in the well-known Kurdjumov–Sachs or Nishiyama–Wasserman orientation relationships. 
From a micromechanical perspective the strain energy minimization led to an important development, which is the phenomenological theory of martensite crystallography \cite{Wechsler1953, Bowles1954}, which is to a large extent in agreement with experimental observations. 
These descriptions are different but related to global energetic considerations, which use the overall energy of the system to predict the formation of microstructures.

In the present work we focus on the long-ranged elastic effects, which are resulting from mismatches between different martensite variants.
They lead to elastic misfits, which induce interactions not only between neighboring phases and grains, but also depend on and influence the entire microstructure.
As discussed in \cite{Sandt2023a}, the proper minimization of the elastic energy requires to keep all these interactions, as an artificial cutoff leads to improper predictions of variant selections.
However, this global perspective introduces an additional degree of complexity to the overall problem, as now a huge space of microstructural configurations needs to be considered.}
For example, for a microstructure consisting of $N$ grains and two martensite variants, we therefore have altogether $2^N$ combinations, and the energy needs to be determined for each of them to find the global ground state, which corresponds to the (low temperature) thermodynamic equilibrium.
It is obvious that this configuration space is too large already for moderate grain size numbers, and therefore not only a brute force energy minimization but also the use of heuristic approaches like simulated annealing are limited and require large amounts of computing times. 
If we consider e.g.~a system consisting of just $N=20$ grains, the number of configurations is $2^N\simeq {\cal O}(10^6)$;
for a typical FFT calculation in 2D with sufficient accuracy the single core runtime on current processors is of the order of one minute.
This implies that scanning all configurations to determine the global energetic minimum requires about 17,000 core hours of computing time.
The question, which arises also generally in the context of materials science, is how this optimization problem can be solved efficiently via quantum computing.

A general quantum computer with a sufficient number of qubits is not available yet, but the technology named quantum annealing (QA)~\cite{Finnila1994,Brooke1999,Kadowaki1998,Morita2008,Rajak2023} with several thousands of qubits and couplers is available.
In contrast to general-purpose quantum computers only specific classes of problems can be handled via QA, i.e. binary quadratic optimization problems~\cite{Warren2018}.
The basic concept of QA is the initialization of its qubits in well defined Hamiltonians, whose ground state is unique and known~\cite{Johnson2011}.
This Hamiltonian is then changed adiabatically under the operation at cryogenic temperatures to the desired, final Hamiltonian, what allows the conversion of the ground state to the final Hamiltonian~\cite{Johnson2011,Boixo2014} and therefore enables efficient global energy minimizations.
The application of QA in materials science is still rare, however, few publications focus on sampling techniques via QA~\cite{Sandt2023b,Nelson2022a,Nelson2022b,Mandra2017,Koenz2019,Mukherjee2019,Yamamoto2020}, phase transitions in the transverse field Ising model~\cite{Harris2018}, critical phenomena in frustrated magnetic systems~\cite{Kairys2020}, energy calculations of defective graphene structures~\cite{Camino2023} and the automated design of metamaterials~\cite{Kitai2020}.
Instead, the focus of actual research concerning quantum annealing rather lies on performance tests and benchmarking of quantum annealing against conventional approaches~\cite{Juenger2021,Parekh2015,Yan2022,King2023}.

In our previous publication \cite{Sandt2023a} we succeeded to map the martensite problem to a quantum annealer formulation, i.e.~a binary quadratic model expressed through an Ising model or equivalently in terms of a quadratic unconstrained binary optimization~\cite{Warren2018}, which allows to find the true ground state configurations even for many thousand grains in an extremely short time~\cite{Sandt2023a}.
This approach builds up on earlier works in \cite{Sherrington2008,Kartha1991,Sethna1992,Vasseur2011}.
Although this conceptual progress demonstrates well the benefits of quantum annealing for materials science modeling, the chosen example was limited in particular by the restriction to just two variants, whereas e.g.~typical shape memory alloys or martensites in steels posses many more variants~\cite{Polatidis2020,Wang2022,Manosa2016,Otsuka2005}.
In the present paper we demonstrate how the approach can be generalized to multi-variant martensites 
and show how the resulting variant distributions are influenced by different tensile strains and random grain rotations.

\section{Methods}

The overall concept of the investigations is that we assume that in a microstructure, as depicted e.g.~in Fig.~\ref{fig2}, each (martensite) grain is allowed to select between different variants, such that the overall elastic energy is minimized in equilibrium.
As the grains are coherently connected to each other, a variant flip as sketched in Fig.~\ref{fig1}, leads possibly to an energetically unfavorable distortion of the surrounding grains, and therefore the discrete optimization of the variants is a nontrivial problem.

To achieve a description in the framework of continuum elasticity, we consider eigenstrain (stress free strain) expressions of the form
\begin{equation} \label{eq1}
\epsilon_{ij}^{(0)}(\mathbf{r}) = \sum_{n=1}^N \theta_n(\mathbf{r}) \sum_{k=1}^K s_{\nu}\epsilon_{ij}^{(0,n,k)}.
\end{equation}
Here, the function $\theta_n(\mathbf{r})$ is equal to one inside grain $n$ and vanishes everywhere else.
The eigenstrain in each grain with index $n$ can be a superposition using $K$ ``spin variables'' $s_{\nu}=s_{n,k}=\pm 1$ {with the total number of spins $N_{\nu}=N\cdot K$}, which weight eigenstrain tensor components $\epsilon_{ij}^{(0,n,k)}$.
Therefore, we have altogether $2^K$ martensite variants in each grain by suitable selection of the parameters $\epsilon_{ij}^{(0,n,k)}$.

To determine the thermodynamic equilibrium state (minimum elastic energy), first the elastic problem has to be solved for a given eigenstrain.
This can be expressed through the minimization of the elastic energy with respect to the displacement field $u_i$ in combination with suitable mechanical boundary conditions.
In case of isotropic linear elasticity with phase and variant independent Lam\'e coefficient and shear modulus, the elastic energy functional reads
\begin{equation}
    E_\mathrm{el} = \int_V \left( \frac{1}{2} \lambda (\epsilon_{kk}-\epsilon_{kk}^0)^2 + \mu (\epsilon_{ik}-\epsilon_{ik}^0)^2 \right) dV
\end{equation}
with the system volume $V$ and the strain tensor $\epsilon_{ik}=\frac{1}{2} ( \partial_i u_k + \partial_k u_i)$.
In the above expression, the Einstein sum convention is used.

Contrary to the optimization of the displacement and strain fields as continuous variables, the variant selection becomes a discrete problem on top of the preceding step.
Following the Fourier transformation approach outlined in \cite{Sandt2023a}, we can calculate the elastic energy $E(\{s_{\nu}\})$ of the microstructure for a given ``spin configuration'' $\{s_{\nu}\}$.
We note that the approach does not necessarily require to use Fourier transformation methods, but e.g.~also finite element methods can be used, provided that the elastic interactions can be computed with sufficient accuracy, as discussed in \cite{Sandt2023a}.

As the expression for the elastic energy is quadratic in the framework of the linear theory of elasticity, the final expression for the elastic energy can be expressed as
\begin{equation} \label{eq2}
    E(\{s_{\nu}\}) = E_0 + \sum_{\nu=1}^{N_{\nu}} \tilde{H}_{\nu} s_{\nu} + \sum_{\nu,\eta=1}^{N_{\nu}} \tilde{J}_{\nu \eta} s_{\nu} s_\eta
\end{equation}
with a quadratic, symmetric matrix $\tilde{J}_{\nu \eta}$.
The coefficients $E_0, \tilde{H}_{\nu}$ and $\tilde{J}_{\nu \eta}$ can be expressed directly in terms of the Fourier transformation solution, or alternatively be calculated using any elastic solver, which delivers the total elastic energy $E(\{s_{\nu}\})$ for a given spin configuration $\{s_{\nu}\}$.
We note that the above form of the energy holds for arbitrary values of the spin variables and does not apply only to the special case $s_{\nu}=\pm 1$.
Therefore, we obtain directly 
\begin{equation}
    E_0 = E(\{s_{\nu}\equiv 0\}),
\end{equation}
where all spins have the value 0, which we also denote as the austenite reference state.
Next, we perform calculations where all spins but one vanish, and the selected one has either the value $+1$ or $-1$.
From that we get
\begin{equation}
E(\{s_i=\pm 1,\mathrm{all\,others\, 0}\}) = E_0 \pm \tilde{H}_i + \tilde{J}_{ii}.
\end{equation}
Therefore, we can extract the diagonal element for the self-interactions as
\begin{eqnarray}
\tilde{J}_{ii} = \frac{1}{2}& \Big[& E(\{s_i=+1,\mathrm{all\,others\, 0}\}) \nonumber \\
&+& E(\{s_i=-1,\mathrm{all\,others\, 0}\} - 2E_0 \Big]
\end{eqnarray}
and consequently
\begin{eqnarray}
    \tilde{H}_i = \frac{1}{2}&\Big[& E(\{s_i=+1,\mathrm{all\,others\, 0}\}) \nonumber \\
    &-& E(\{s_i=-1,\mathrm{all\,others\, 0}\})\Big].
\end{eqnarray}
Finally, to calculate the interaction coefficients $\tilde{J}_{ij}$ for $i\neq j$ we perform elastic calculations with two nonvanishing spins $s_i=s_j=1$.
With
\begin{equation}
  E(\{s_i=s_j= 1,\mathrm{others\, 0}\}) = E_0 + \tilde{H}_i + \tilde{H}_j + 2\tilde{J}_{ij} + \tilde{J}_{ii} + \tilde{J}_{jj},  
\end{equation}
where we can directly calculate $\tilde{J}_{ij}$ using the expressions obtained above
\begin{eqnarray}
 2\tilde{J}_{ij} &=& E(\{s_i=s_j=+1,\mathrm{all\,others\, 0}\}) \nonumber \\
 &-& E(\{s_i=+1,\mathrm{all\,others\, 0}\}) \nonumber \\
 &-& E(\{s_j=+1,\mathrm{all\,others\, 0}\}) + E_0.
\end{eqnarray}
For zero stress boundary conditions, we get for the eigenstrain of the type (\ref{eq1}) the simplification $E_0=0$ and $\tilde{H}_i=0$ for all $i$.

For the use of the quantum annealer, we restrict the allowed spin values to $s_\nu=\pm 1$ and we also have to rewrite the expression (\ref{eq2}) in a slightly different way, which excludes in particular self interactions of the type $s_is_i$.
The Hamiltonian reads
\begin{equation} \label{QAIsing}
H = H_0 + \sum_{i=1}^{N_{\nu}} H_i s_i + \sum_{i=1}^{N_{\nu}}\sum_{j=i+1}^{N_{\nu}} J_{ij} s_i s_j,
\end{equation}
where the offset $H_0$ only affects the energy value but not the spin configuration minimizing the energy.
Comparison with the expression (\ref{eq2}) above gives 
\begin{eqnarray}
    H_0 &=& E_0 + \sum_{i=1}^{N_{\nu}} \tilde{J}_{ii}, \\
    H_i &=& \tilde{H}_i, \\
    J_{ij} &=& 2 \tilde{J}_{ij} \quad \mbox{for}\,\,i<j.
\end{eqnarray}
This general formulation has the advantage that it decouples the approach for setting up the Ising coefficients for the annealer from the specific linear elastic energy solver. 
Nevertheless, one has to keep in mind that the formulation (\ref{eq2}) relies on the assumption of equal elastic constants in all phases and variants and coherent boundary conditions.
However, it is valid also in situation beyond the original work \cite{Sandt2023a}, where we used only a single spin to discriminate between the two martensite variants in each grain.
Here, the same approach also applies for multiple variants using several spins per grain.

On the quantum annealer level the determined Ising coefficients are used to define an energy landscape, where superconducting loops with clockwise or anticlockwise circulating currents define qubits with different spin states~\cite{Johnson2011}.
These superconducting loops interact with external flux biases, which allow to control energy difference and barrier height of the constructed energy landscape~\cite{Johnson2011}.
Starting point of the annealing process is the initialization of the system in the ground state of a known Hamiltonian $H_0\sim -\sum_i \sigma_i^x$ where $\sigma_i$ denotes the Pauli matrices, i.e. corresponding to a strong transverse magnetic field~\cite{Boixo2014,Ronnow2014}.
This Hamiltonian is turned during the annealing into the desired Ising model (\ref{QAIsing}) for which an energetic minimum is sought, $\mathrm{min}_{\{s_i=\pm 1\}}H_p$ ~\cite{Warren2018}.
The Hamiltonians $H_0$ and $H_p$ do not commute~\cite{Warren2018}, and the time taken by the initial Hamiltonian to reach the low energy state is sufficiently large to establish the validity of the adiabatic theorem of quantum mechanics~\cite{Lucas2014}, which postulates that a system stays in its eigenstate during adiabatic changes.
In contrast to classical approaches, further quantum mechanical principles like tunneling to leave metastable regions and the usage of entangled states inside quantum annealing processors (QPU) are employed~\cite{Lanting2014}.
Especially if energetically close low energy states exist, the QA process does not always determine the true ground state, therefore a suitable number of repetitions is performed and the lowest detected energy is chosen.
Additionally, for larger systems, hybrid quantum annealing utilises classical algorithms and the interplay with quantum annealing to address areas with high computational demands using a QPU coprocessor working with generic parameters for up to 11616 spin variables on the D-Wave Advantage system~\cite{Raymond2023,Berwald2019, Leap}.

\section{Results}

\begin{figure}
\begin{center}
    \includegraphics[width=0.5\textwidth, trim=10cm 1cm 8cm 2cm, clip=true]{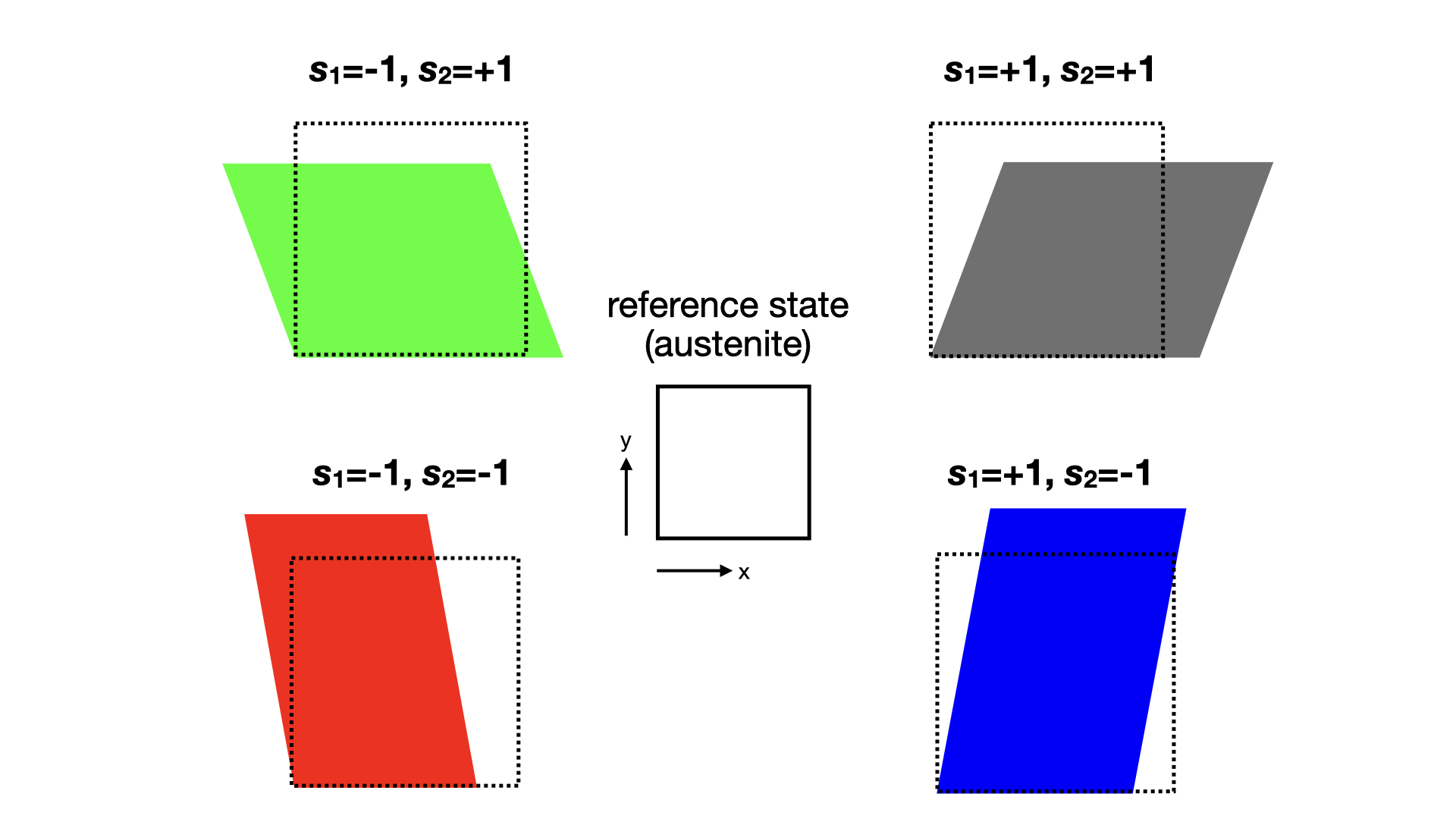}
    \caption{Visualization of the variants used in the present study. Spin $s_1$ corresponds to the shear deformation, $s_2$ to the tetragonal deformation, see Eq.~(\ref{eq3}).}
    \label{fig1}
\end{center}
\end{figure}

To illustrate the workflow, we consider a case with four variants per grain, i.e.~two spins per grain.
Similar to the previous work \cite{Sandt2023a} we use specific cases of shear and tetragonal distortions, which are now combined and lead to eigenstrains
\begin{equation} \label{eq3}
\epsilon_{ij}^{(0,n)} = s_{n,1} 
\left(
\begin{array}{ccc}
0 & \epsilon_0 & 0 \\
\epsilon_0 & 0 & 0 \\
0 & 0 & 0
\end{array}
\right)
+ s_{n,2}
\left(
\begin{array}{ccc}
\epsilon_1 & 0 & 0 \\
0 & -\epsilon_1 & 0 \\
0 & 0 & \epsilon_1
\end{array}
\right),
\end{equation}
where $n$ numerates the grains. 
For the first application we assume that the individual tensors of the shear (involving the strain parameter $\epsilon_0$) and tetragonal deformations (associated with $\epsilon_1$) are the same in all grains;
later we will also discuss the case of grain rotations, where the tensors differ from grain to grain.
Fig.~\ref{fig1} visualizes the different strain variants and illustrates the color codings used for the following plots.
The grain structure is generated using a Voronoi tesselation in two dimensions, and we assume plane strain conditions, using isotropic linear theory of elasticity with periodic boundary conditions.
According to the above approach, first the self energies related to each spin are calculated, followed by a computation of spin-spin interactions, from which the Ising vector components $H_n$ and interaction parameters $J_{ij}$ are extracted.

Apart from the self generated internal stresses due to the different variants, also a homogeneous external (average) strain field $\langle{\epsilon}_{ij}\rangle$ can be superimposed, which leads to an additional contribution to the magnetic field vector $H_n$, which couples the external strain $\langle\epsilon_{ij}\rangle$ to the $\mathbf{k}=0$ mode of the Fourier transform of the eigenstrain tensor.
This means that a change of the external strain leaves the coupling constants $J_{ij}$ invariant and only modifies a contribution to $H_n$, which can be treated analytically, therefore minimizing the computational cost for repeated calculations with different external boundary conditions.

\begin{figure}
\begin{center}
\includegraphics[width=0.5\textwidth, trim=3cm 2cm 1cm 0cm, clip=true]{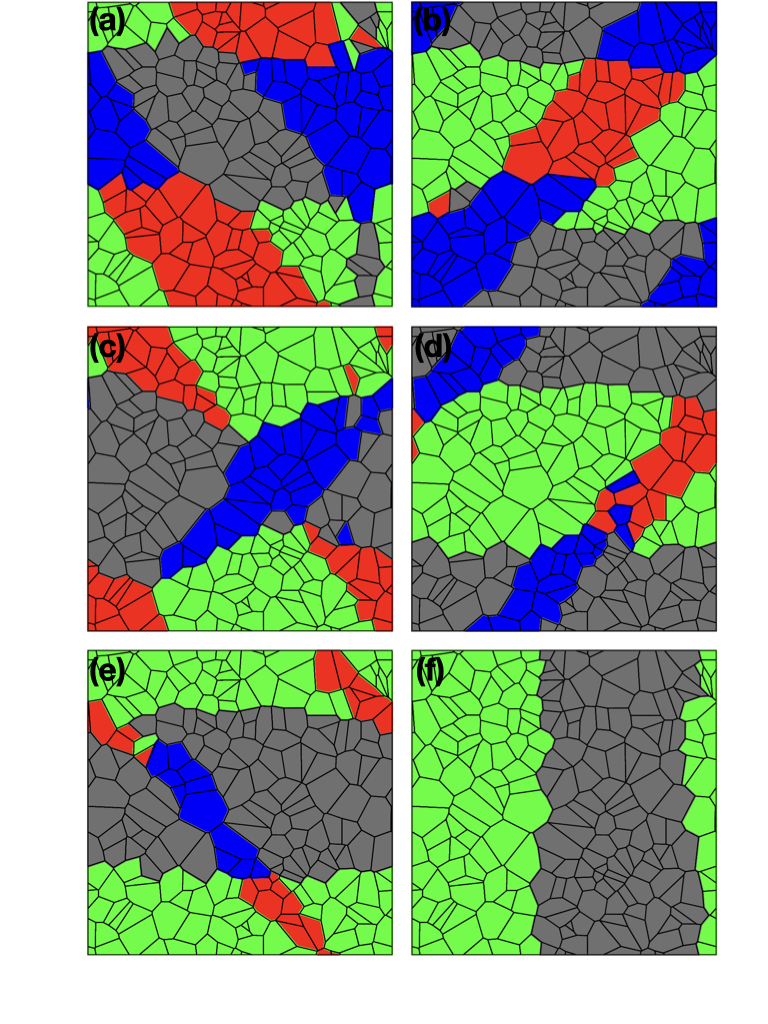}
\caption{Equilibrium variant distribution with uniform grain orientation. The microstructure consist of 200 grains and a tensile strain is applied in horizontal ($x$) direction.
The tensile strain is \textbf{(a)} $\langle\epsilon_{xx}\rangle/\epsilon_1 = 0$,  \textbf{(b)} $\langle\epsilon_{xx}\rangle/\epsilon_1 = 0.25$, \textbf{(c)} $\langle\epsilon_{xx}\rangle/\epsilon_1 = 0.5$, \textbf{(d)} $\langle\epsilon_{xx}\rangle/\epsilon_1 = 0.75$, \textbf{(e)} $\langle\epsilon_{xx}\rangle/\epsilon_1 = 1.0$ and \textbf{(f)} $\langle\epsilon_{xx}\rangle/\epsilon_1 = 1.5$.
In all figures $\epsilon_0=\epsilon_1$ and the Poisson ratio is chosen as $\nu=1/4$ (i.e. $\lambda=\mu$).
}
\label{fig2}
\end{center}
\end{figure}
Fig.~\ref{fig2} shows a sequence of equilibrium microstructures and their response to a tensile strain in horizontal ($x$) direction for $\epsilon_1=\epsilon_0$.
Here, the computer generated, random 2D microstructure consists of 200 grains.
Each grain selects its optimum variant state $(s_{n,1}, s_{n,2})$ to minimize the total energy.
In panel (a) the fixed external strain vanishes, {which implies that in the ground state the volume fractions of the blue and green (resp.~red and grey) variants should be close to each other, as then the positive and negative contributions to the shear and tetragonal (eigen)strain cancel each other to minimize the elastic energy.
The arrangement of the differently coloured patches is then a result of the mutual spin-spin interactions. 
We can observe in panel (a) a structure of horizontal stripes with alternating layers of shear strain (blue-grey and red-green).
Such stripes with (100) orientation have been discussed already in \cite{Sandt2023a}, as the formation of these structures minimizes the overall shear deformation.
Additionally, tilted stripe pairs red-blue and and green-grey correspond to the same tetragonal strain and annihilating shear strain, similar to the patterns for pure tetragonal eigenstrain discussed in \cite{Sandt2023a}.

The patterns formed by several variants can be better understood by analyzing the elastic anisotropy of the transformation. Specifically, the normal to the interface between two variants must be close to a direction that minimizes the elastic kernel of the transformation from one variant to the neighboring one. This can be demonstrated for plate-like domains \cite{Khachaturyan2008}, but is also very commonly observed in complex microstructures whose anisotropy is of elastic origin, such as cuboidal microstructures in superalloys \cite{Wang1998}, Widmanst\"atten structures \cite{Lebbad2021} or chessboard structures in Co-Pt alloys \cite{Yann1998}. 
The elastic kernel can be obtained analytically in Fourier space \cite{Khachaturyan2008} or can be calculated in real space, using a regular stripe pattern of both variants. Indeed, in this geometry, only Fourier modes perpendicular to the stripes remain, and the elastic energy is proportional to the elastic kernel. In the following, the orientation dependence of the pattern is analyzed by considering successively each pair of variants.
 We assume fixed vanishing average strain and equal volume fractions of the two involved variants.
The angular dependence of relevant combinations of stripe pairs is shown in Fig.~\ref{orientation::fig}.
\begin{figure*}
\begin{center}
\includegraphics[width=0.8\textwidth, trim=19cm 0cm 19cm 0cm, clip=true]{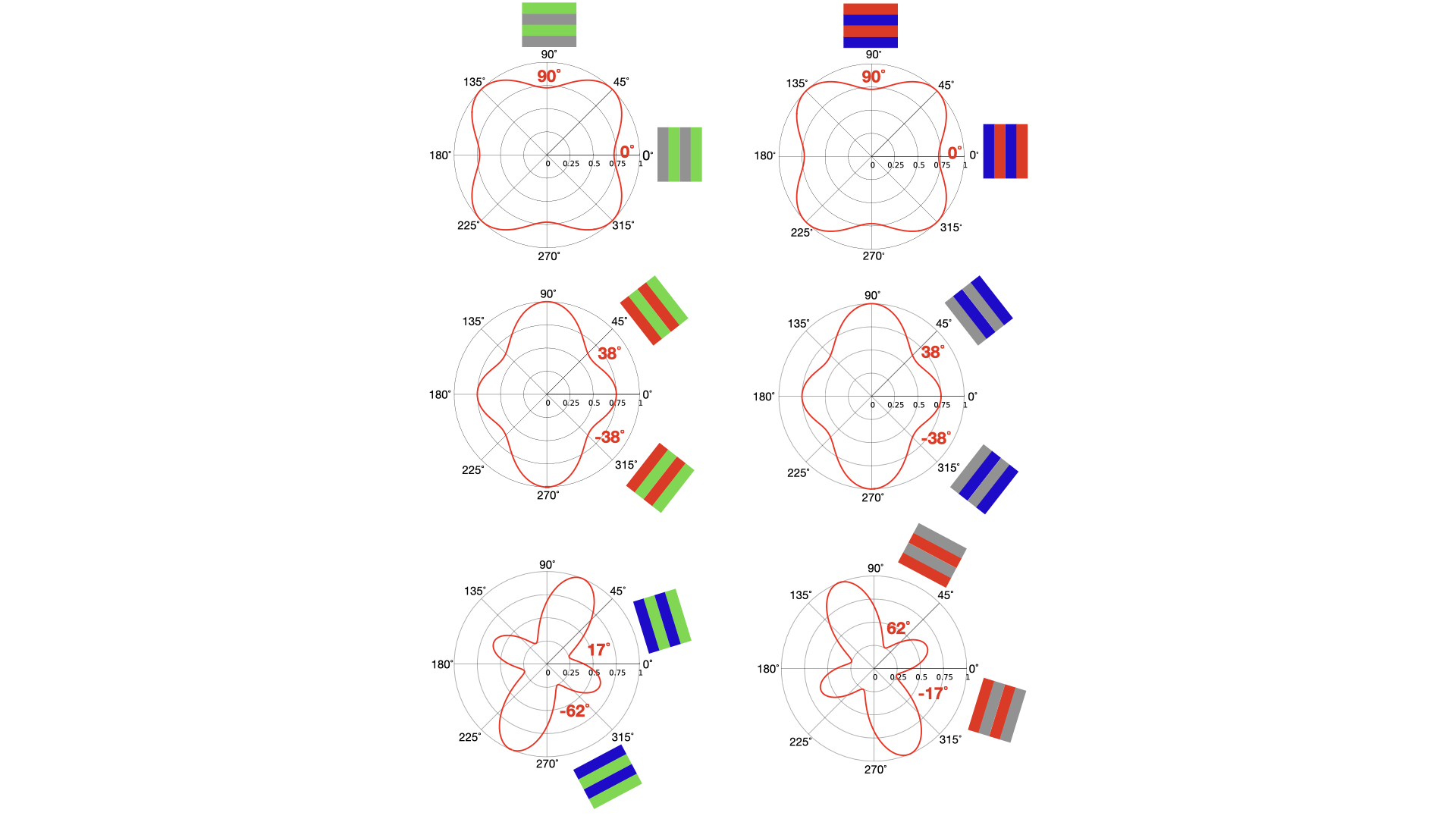}
\caption{Orientation dependence of the normalized elastic energy for equal volume fraction stripes consisting of pairs of variants for $\langle\epsilon_{ij}\rangle=0$ and $\epsilon_1=\epsilon_0$.}
\label{orientation::fig}
\end{center}
\end{figure*}
In these pole figures the angle $0^\circ$ corresponds to vertical stripes, and positive angles to anticlockwise rotations of the pattern.
From this figure, one can conclude that both green-grey and red-blue interfaces are preferred either at $0^\circ$ or $90^\circ$, and this is in agreement with the observed interface orientation between these variant pairs in Fig.~\ref{fig2}(a).

Both green-red and grey-blue pairs have preferred orientations around $\pm 38^\circ$.
Due to constraints of the chosen system sizes and the used periodic boundary conditions, typically deviations from the analytical prediction for infinite system sizes are expected, as discussed in detail in \cite{Sandt2023a}.
In the equilibrium microstructure in Fig.~\ref{fig2}(a) we indeed find inclined interfaces between green-red and grey-blue pairs, which are in line with the theoretical expectation.

Finally, green-blue and red-grey interfaces show a very pronounced anisotropy with a strong cusp at either $17^\circ$ and $-62^\circ$ (green-blue) or $62^\circ$ and $-17^\circ$ (red-grey).
Due to this strong orientation dependence these interfaces either have to appear at these preferred angles or they are strongly supressed.
In fact, the corresponding interface lengths in Fig.~\ref{fig2}(a) are much shorter than the other interfaces.

We note for the subsequent panels of Fig.~\ref{fig2} that the elaborated anisotropy of the interactions is not affected by a homogeneous external strain, and therefore the polar plots in Fig.~\ref{orientation::fig} remain valid also in these cases.
In these pictures, the applied tensile external strain in horizontal direction is incrementally increased.
This favors the variants with a positive ``tetragonal spin value'', and consequently the fraction of green and grey grains increases.
They form more and more pronounced bands in the $(100)$ directions, as explained above, partly interrupted by disappearing inclined red/blue bands with unfavorable tetragonal strain contribution.
Finally, only grains with $s_{n,2}=+1$ remain, and the volume fractions with positive and negative shear are essentially the same, as much as the irregular grain sizes permit.

{
It is useful to emphasize the periodicity of the patterns by looking at a $2\times 2$ supercell of the simulation domain, as shown in Fig.~\ref{fig2supercell}.
\begin{figure*}
\begin{center}
\includegraphics[width=1\textwidth, trim=14cm 0cm 2cm 0cm, clip=true]{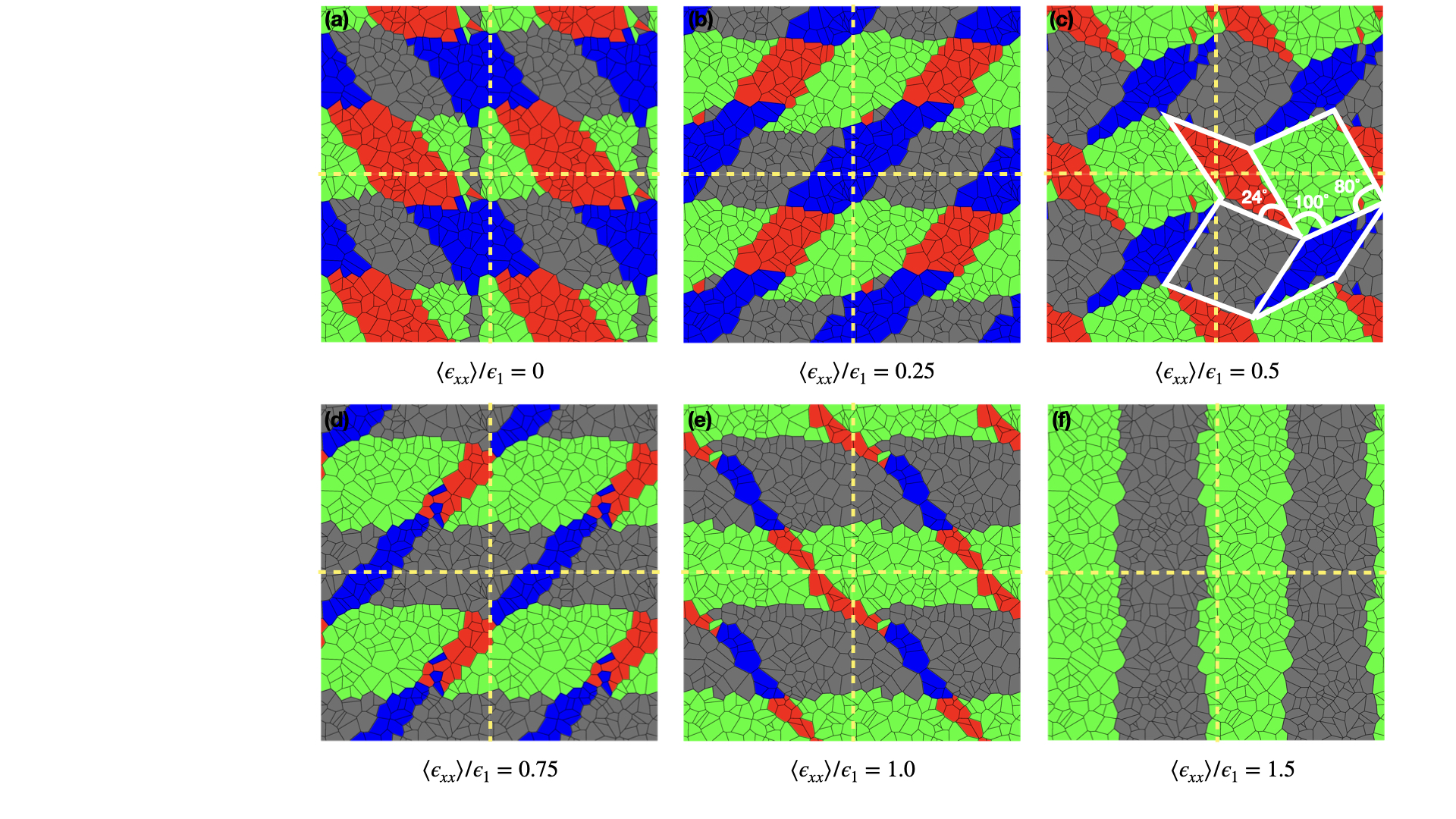}
\caption{$2\times 2$ supercell representation of Fig.~\ref{fig2} to emphasize the microstructure patterns. The dashed yellow lines mark the boundaries or the simulation with periodic boundary conditions. In all figures $\epsilon_1=\epsilon_0$ and $\nu=1/4$.}
\label{fig2supercell}
\end{center}
\end{figure*}
The first striking observation is that the pattern in panel (c) for $\langle \epsilon_{xx}\rangle/\epsilon_0=0.5$ differs significantly from the other cases.
These latter cases show both horizontal and tilted stripes, as analyzed above.
There, grey-green and red-blue interfaces appear at $0^\circ$ or $90^\circ$, and green-blue as well as red-grey interfaces play at most a minor role, as discussed above.
The red-green and grey-blue interfaces are roughly at $\pm 45^\circ$ as a compromise between the preferred analytical predictions for infinite systems and the constraints by the periodic boundary conditions.

The special case $\langle \epsilon_{xx}\rangle/\epsilon_0=0.5$ shows a remarkable topological change, as now interfaces between green-blue, green-red, grey-blue and grey-red appear, whereas the previously important red-blue and green-grey interfaces are absent.
Moreover, the newly appearing interfaces are at their preferred orientations. 
For example, the marked $24^\circ$ angle inside the red lozenge is the difference between the $62^\circ$ minimum energy orientation for the red-grey interface and the $38^\circ$ orientation for the red-green pattern.
This leads to the geometry sketched in  Fig.~\ref{fig2supercell}(c), from which the overall periodic pattern is constructed.
According to the equilibrium angles, this demands that the joint volume fraction of the diamond-shaped green and grey domains is $\sin(80^\circ)/[\sin(24^\circ)+\sin(80^\circ)]\approx 0.7$, and therefore this exceptional pattern is only observed close to a specific external strain value.
We note that this chessboard structure is very similar to the one observed in Co-Pt and (CuAu)$_{1-x}$-Pt$_x$ systems \cite{Yann1998}. 
However, the symmetry is lower in the present case because the green and grey regions are not tetragonal.
}

The same microstructure as in Fig.~\ref{fig2} is subjected to a shear strain $\langle \epsilon_{xy}\rangle$ in (110) direction, and the results of the energy minimization are shown in Fig.~\ref{fig3}.
\begin{figure*}
\begin{center}
\includegraphics[width=0.8\textwidth, trim=7.5cm 0cm 3cm 0cm, clip=true]{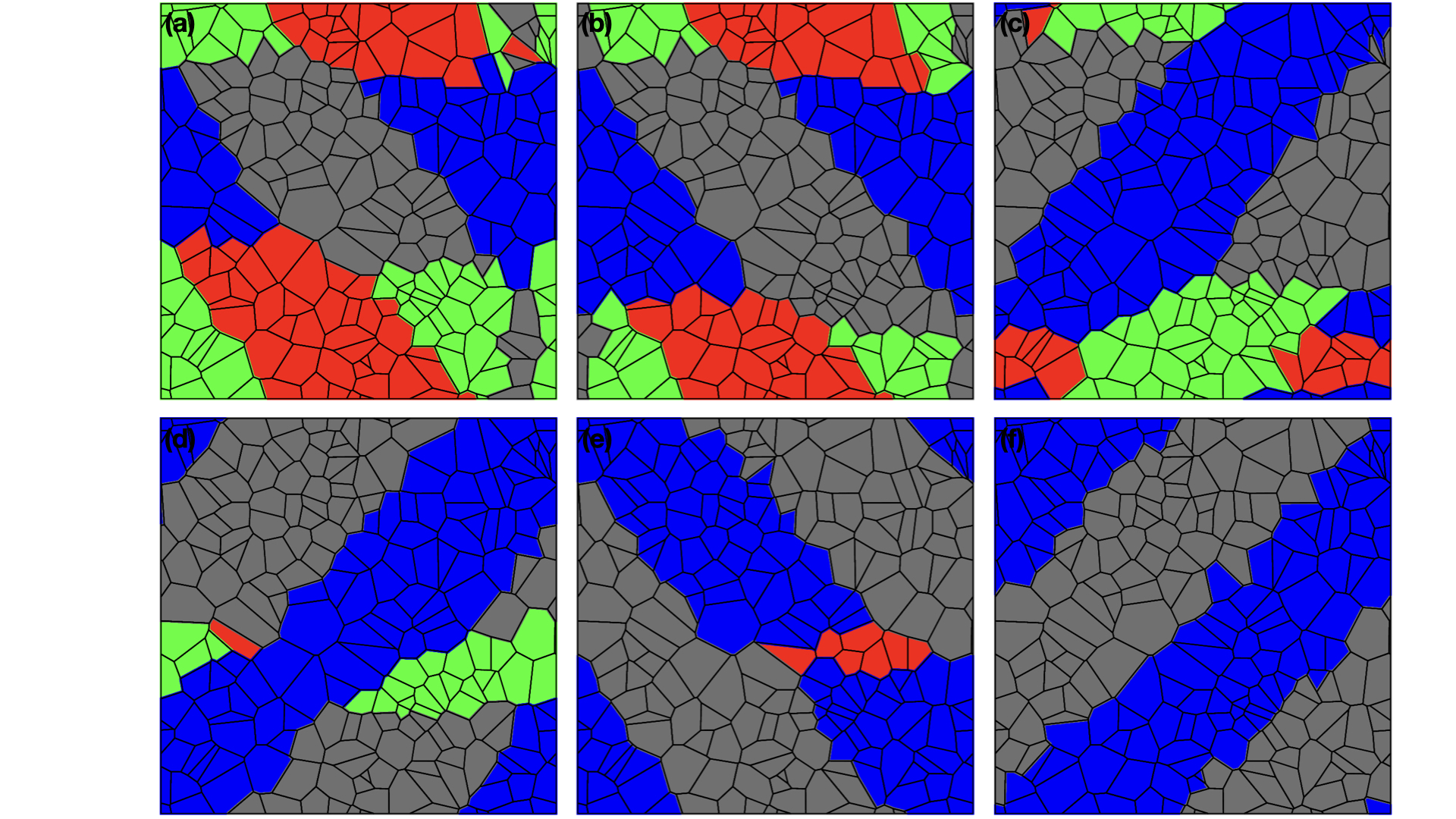}
\caption{Equilibrium variant distribution with uniform grain orientation. The microstructure consist of 200 grains and an average shear strain $\langle \epsilon_{xy}\rangle$ is applied.
The strain is \textbf{(a)} $\langle\epsilon_{xy}\rangle/\epsilon_0 = 0$,  \textbf{(b)} $\langle\epsilon_{xy}\rangle/\epsilon_0 = 0.25$, \textbf{(c)} $\langle\epsilon_{xy}\rangle/\epsilon_0 = 0.5$, \textbf{(d)} $\langle\epsilon_{xy}\rangle/\epsilon_0 = 0.75$, \textbf{(e)} $\langle\epsilon_{xy}\rangle/\epsilon_0 = 1.0$ and \textbf{(f)} $\langle\epsilon_{xy}\rangle/\epsilon_0 = 1.5$.
In all figures $\epsilon_0=\epsilon_1$, and the chosen Poisson ratio is $\nu=1/4$ (i.e. $\lambda=\mu$).
}
\label{fig3}
\end{center}
\end{figure*}
The starting configuration without external strain (panel (a)) is therefore identical to Fig.~\ref{fig2}(a).
An increase of the shear strain favors variants with positive $s_{n,1}$, as they accommodate the given external strain to a large amount.
Therefore, for large positive shear strain $\langle\epsilon_{xy}\rangle/\epsilon_0$ values only grey and blue patches remain. 
As they differ by the tetragonal eigenstrain only, they align as inclined bands to minimize the total elastic energy, as discussed in detail in \cite{Sandt2023a}.
{
In particular, grey-green and red-blue interfaces are at $0^\circ$ or $90^\circ$, as predicted in Fig.~\ref{orientation::fig}.
Similarly, blue-green and red-grey interfaces have orientations close to the expected ones ($\pm17^\circ,\pm62^\circ$).
For blue-grey and red-green interfaces the periodicity constraint enforces orientations close to $45^\circ$.

\begin{figure}
\begin{center}
\includegraphics[width=0.45\textwidth, trim=18cm 2cm 13cm 1.5cm, clip=true]{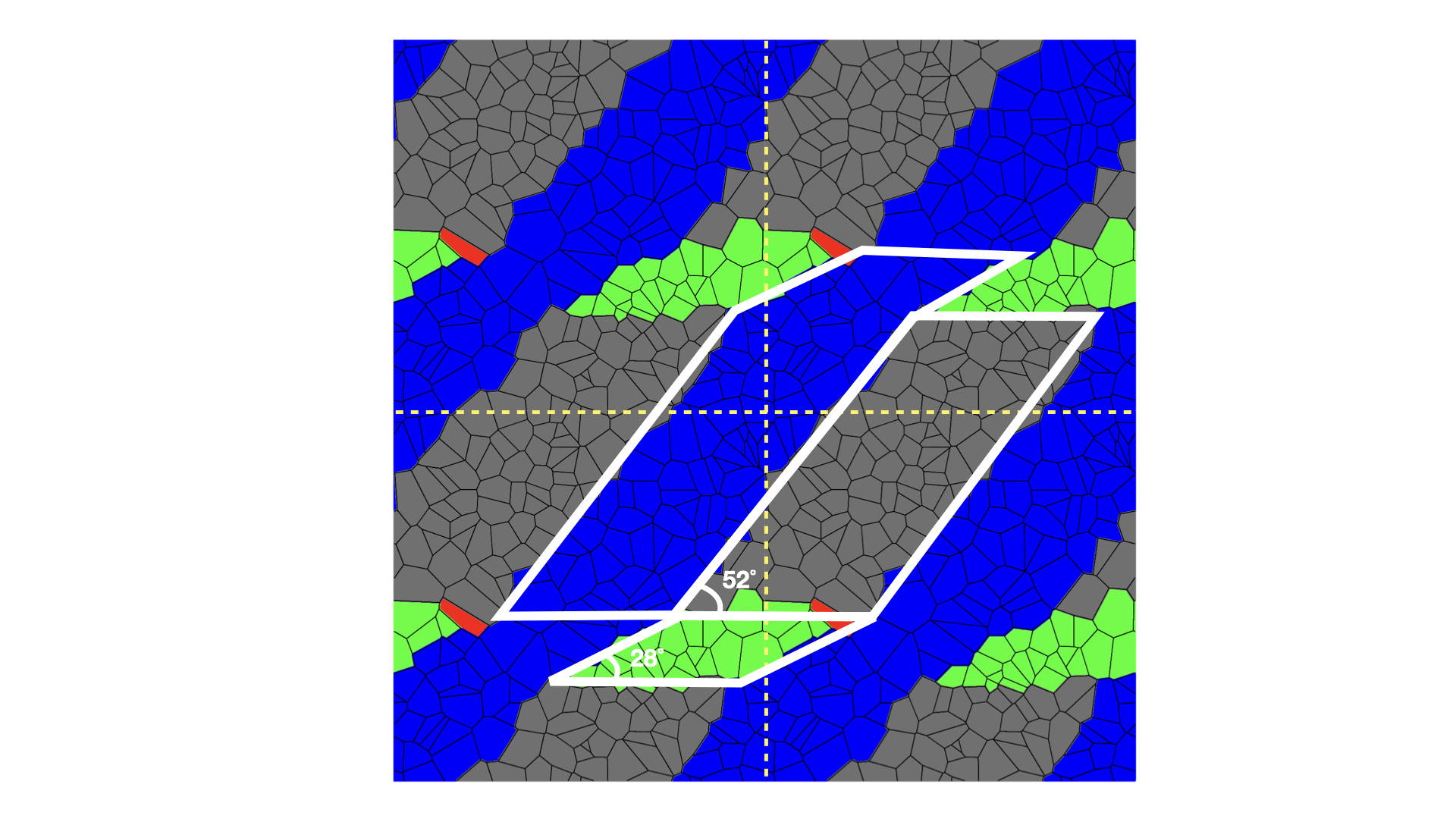}
\caption{$2\times 2$ supercell of the simulation domain for external shear $\langle\epsilon_{xy}\rangle/\epsilon_0=0.75$ from Fig.~\ref{fig3}(d) with $\epsilon_1=\epsilon_0$.
    Apart from a negligible amount of the red variant three phases are present here.
    The sketch shows the geometrical explanation of the equilibrium pattern.}
\label{fig:3domains}
\end{center}
\end{figure}

An interesting case is the three-variant configuration in Fig.~\ref{fig3}(d) (apart from a negligible fraction of the red variant).
A $2\times 2$ supercell of the equilibrium microstructure is shown in Fig.~\ref{fig:3domains}.
The overall $45^\circ$ alignment is imposed by periodic boundary conditions.
The green domain slightly changes the orientation of the blue-grey interfaces so that they are rotated closer to the optimal orientation $90^\circ-38^\circ=52^\circ$.
This configuration would not be favorable at $-45^\circ$ (instead of the $45^\circ$) because the blue-grey interfaces would be rotated away from the optimal orientation. 
Instead, for $-45^\circ$, red domains appear in the blue platelets (see Fig.~\ref{fig3}(e)).
}

For the present four variant setup, the response of the microstructure does not only depend on the ratio $\langle \epsilon_{ij}\rangle/\epsilon_0$, but also on the relative relevance of shear and tetragonal deformations, as expressed through the dimensionless parameter $\epsilon_1/\epsilon_0$.
Whereas in the above investigations we have fixed this ratio to $\epsilon_1/\epsilon_0=1$, Fig.~\ref{fig:e0_e1} shows the influence of this parameter for fixed external strain.
Here, we note that a change of $\epsilon_0\to a \epsilon_0$ and $\epsilon_1\to b \epsilon_1$ does not require to repeat the entire computations for the interaction parameters $J_{ij}$ and the external magnetic field $H_i$.
Instead, it follows from the theory of linear elasticity that the coupling constants $J_{ij}$ are quadratic in the rescaling factors $a,b$, and $H_i$ scales linearly with them.
In detail, for a spin pair $(i, j)$, where both of them refer to shear transformations, the coupling constants are rescaled according to $J_{ij}\to a^2 J_{ij}$, and similarly for two ``tetragonal spins'' as $J_{ij}\to b^2 J_{ij}$.
For mixed interactions between a tetragonal and a shear degree of freedom, the rescaling obeys $J_{ij}\to ab J_{ij}$.
The ``magnetic'' terms $H_n$ scale linearly with either $a$ or $b$, depending on the spin type.

\begin{figure*}
\begin{center}
\includegraphics[width=0.8\textwidth, trim=7.5cm 19.5cm 3cm 0cm, clip=true]{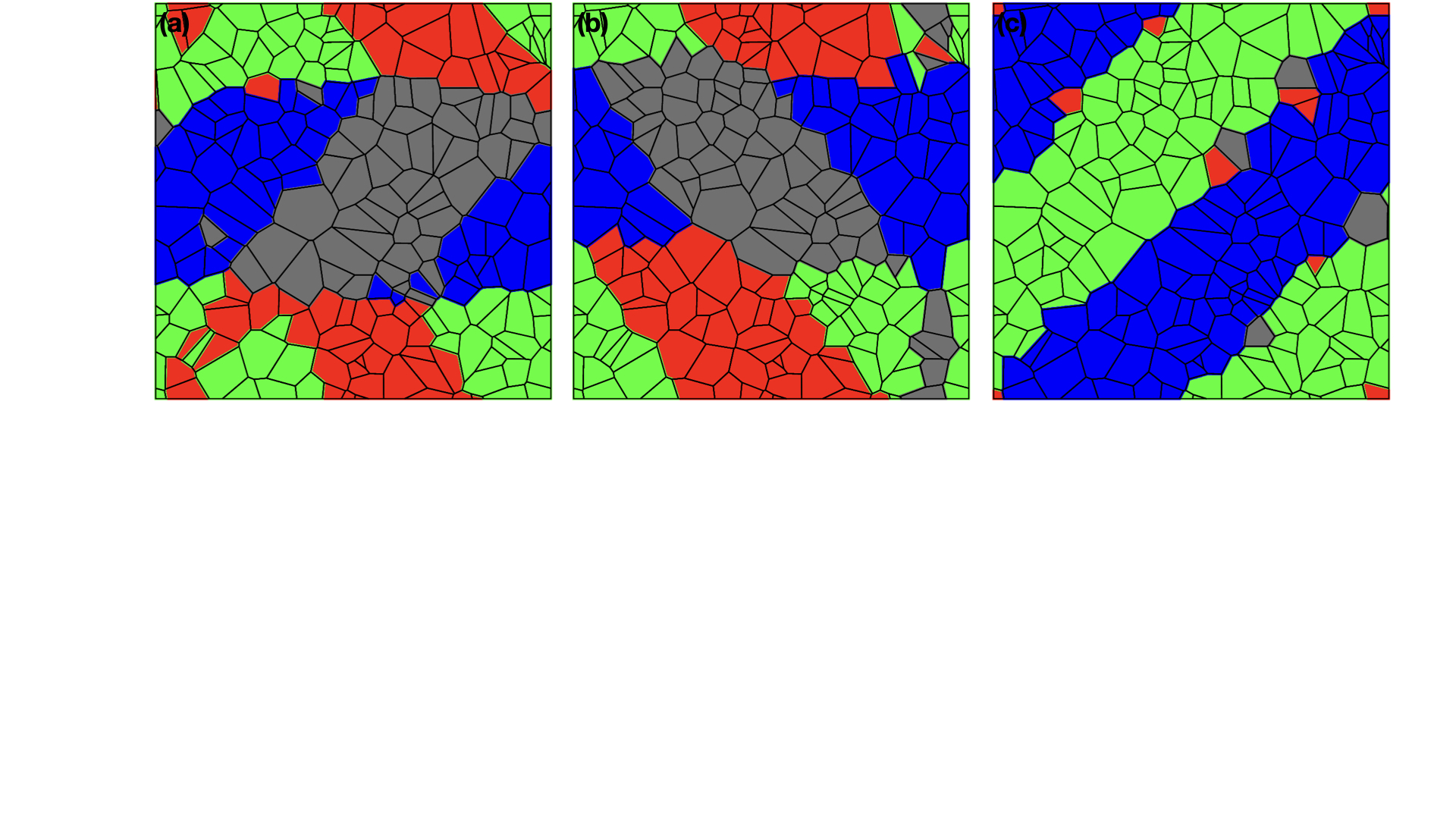}
\caption{Equilibrium variant distribution with uniform grain orientation  for $\langle\epsilon_{ij}\rangle=0$. The microstructure consist of 200 grains and the dimensionless parameter $\epsilon_1/\epsilon_0$ is varied.
The ratios are \textbf{(a)} $\epsilon_1/\epsilon_0 = 0.2$,  \textbf{(b)} $\epsilon_1/\epsilon_0 = 1$ and \textbf{(c)} $\epsilon_1/\epsilon_0 = 5$.
In all figures the chosen Poisson ratio is $\nu=1/4$ (i.e. $\lambda=\mu$).
}
\label{fig:e0_e1}
\end{center}
\end{figure*}

The starting case (center panel in Fig.~\ref{fig:e0_e1}) is again the already discussed reference situation with $\epsilon_1=\epsilon_0$.
In all three pictures, the boundary conditions are fixed to vanishing average strain, $\langle\epsilon_{ij}\rangle=0$.
The right panel shows the equilibrium microstructure for $\epsilon_1/\epsilon_0=5$, which means that the weight of the tetragonal deformations is significantly larger than that of the shear.
Here, the variants select essentially diagonal bands, as discussed in \cite{Sandt2023a}, in order to match the periodic boundary conditions.
The blue and green variants have cancelling tetragonal and shear eigenstrains, such that for equal volume fractions the zero strain boundary conditions are met best.
We can expect that also the opposite colour pair (red-grey) can lead to an energetically equivalent solution, and therefore a weak, locally ``noisy'' pattern, which mixes these two configurations, may appear.
We expect this effect to be more pronounced in larger systems.
One may expect that the opposite case with $\epsilon_1/\epsilon_0=0.2$, which is dominated by the shear eigenstrain, should similarly lead to bands with alternating shear in $\langle100\rangle$ direction, as discussed for the two-variant case in \cite{Sandt2023a}.
However, as can be seen from the left panel of Fig.~\ref{fig:e0_e1}, this is not the case.
Instead, each horizontal stripe consists of blue and grey (or red and green) patches, which share the same shear eigenstrain but have opposite tetragonal strain.
The reason for this assymetry --- as compared to the tetragonal dominated case discussed above --- is that a small, but nonvanishing tetragonal eigenstrain leads also to a tangential mismatch strain at the interfaces between the $\langle100\rangle$ bands.
Therefore, a purely blue horizontal band would lead to a contraction in tangential direction, whereas the compensating green horizontal band would expand in the same direction, hence leading to large mismatch coherency stresses.
Therefore, it is energetically favorable to generate $\langle 100\rangle$ structures with an additional zigzag substructure, such that all four variants are present with equal volume fraction, and only the irregularity of the microstructure adds some minor noise.

{
It is again instructive to look at the angular dependence of the elastic energy for infinite stripe patterns (again for zero mean strain boundary conditions), as it depends on the ratio $\epsilon_1/\epsilon_0$, see Fig.~\ref{epsratiopolar}.
\begin{figure*}
\begin{center}
\includegraphics[width=0.93\textwidth, trim=17cm 0cm 18.5cm 0cm, clip=true]{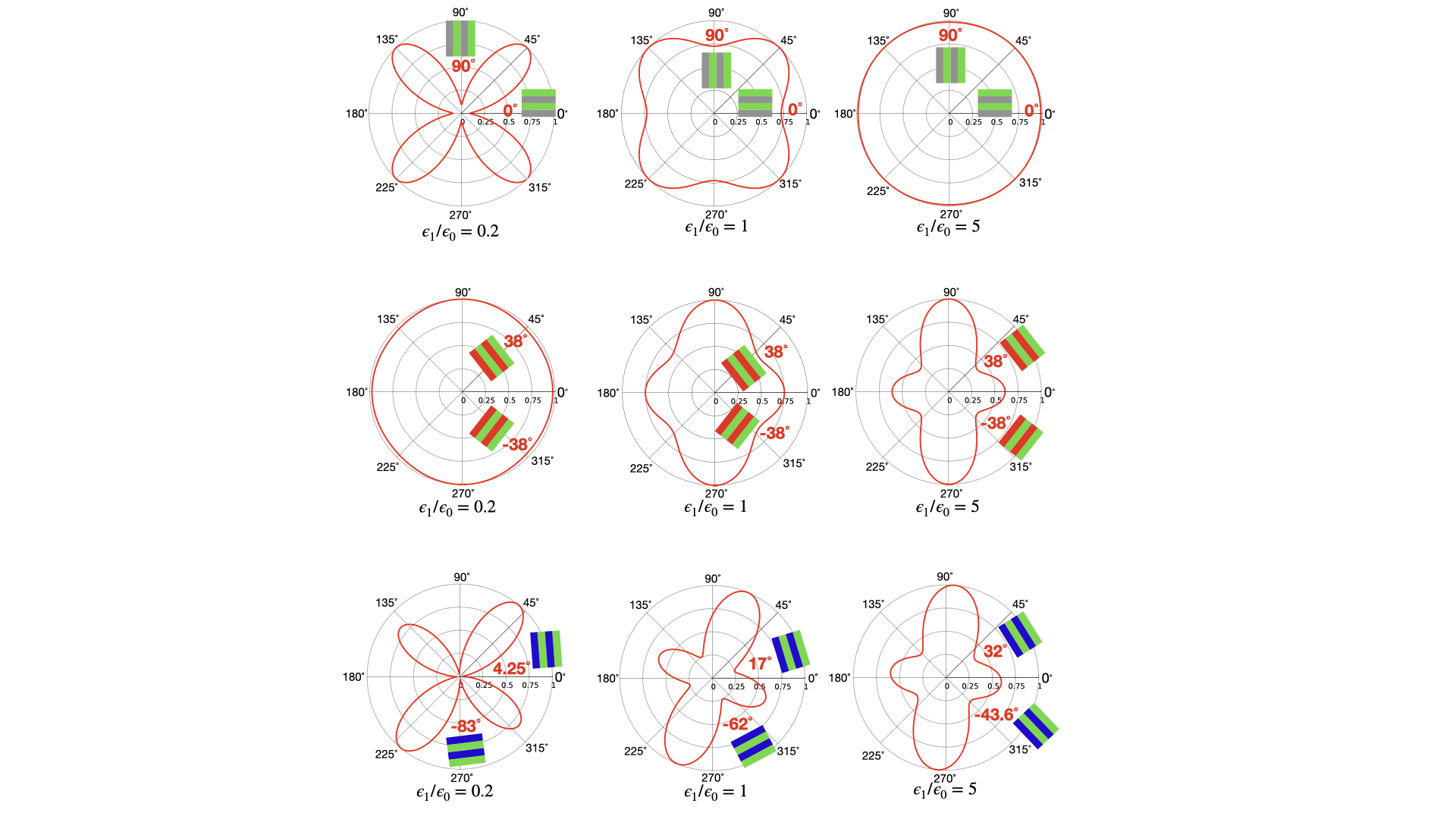}
\caption{Polar representation of the orientation dependence of the elastic energy (scaled to the maximum value) for mismatching pairs of variants.
    The calculations are for fixed strain boundary conditions $\langle\epsilon_{ij}\rangle = 0$. 
    }
\label{epsratiopolar}
\end{center}
\end{figure*}
Here, one can clearly see that the interaction becomes very anisotropic for green-grey interfaces at low ratios $\epsilon_1/\epsilon_0$ and essentially isotropic at high values, with minimum energetic cost at $0^\circ$ and $90^\circ$.
For the green-red interfaces, the trend is opposite.
Green-red (and similarly grey-blue) interfaces have preferred interface orientations $\pm 38^\circ$.
The green-blue and similarly grey-red interfaces have the strongest dependence on $\epsilon_1/\epsilon_0$ and therefore the overall energy minimization clearly favors the optimal orientations for these interfaces.
From this perspective, Fig.~\ref{fig:e0_e1}(a) is particularly interesting, and its zigzag structure becomes obvious in the $2\times 2$ supercell representation in Fig.~\ref{e1e0zigzag}.
\begin{figure}
\begin{center}
\includegraphics[width=0.45\textwidth, trim=18cm 2cm 15cm 1.5cm, clip=true]{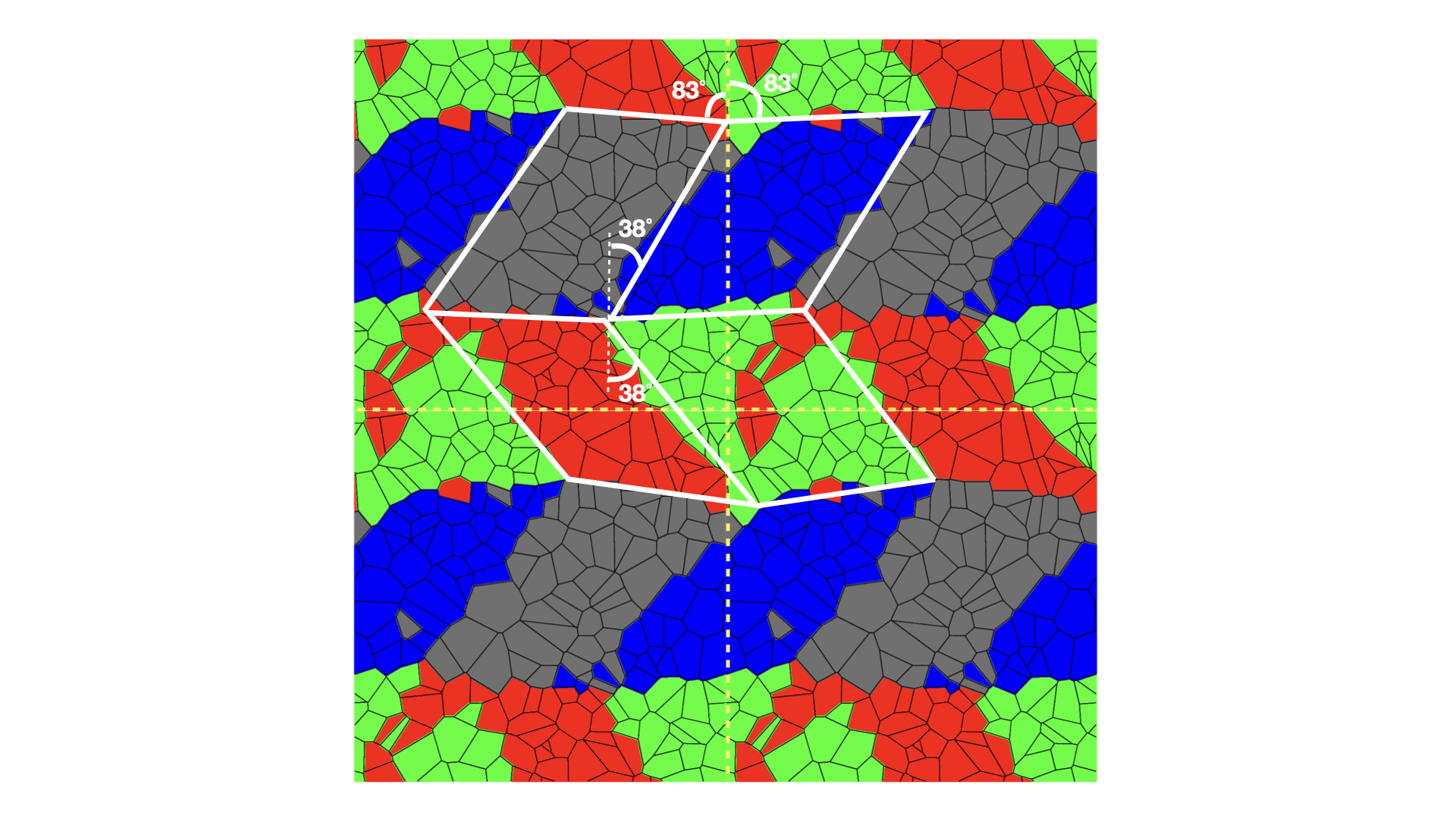}
\caption{$2\times 2$ supercell of the periodic solution Fig.~\ref{fig:e0_e1}(a) with $\epsilon_1/\epsilon_0=0.2$.
    The structure's geometry is sketched and highlights the interface orientations between the variants.}
\label{e1e0zigzag}
\end{center}
\end{figure}
The white lines emphasize the structure and illustrate that for this microstructure all interface orientations are optimal and not affected by periodicity constraints. 
}

\begin{figure*}
\begin{center}
\includegraphics[width=0.9\textwidth, trim=2.5cm 18.0cm 3cm 0cm, clip=true]{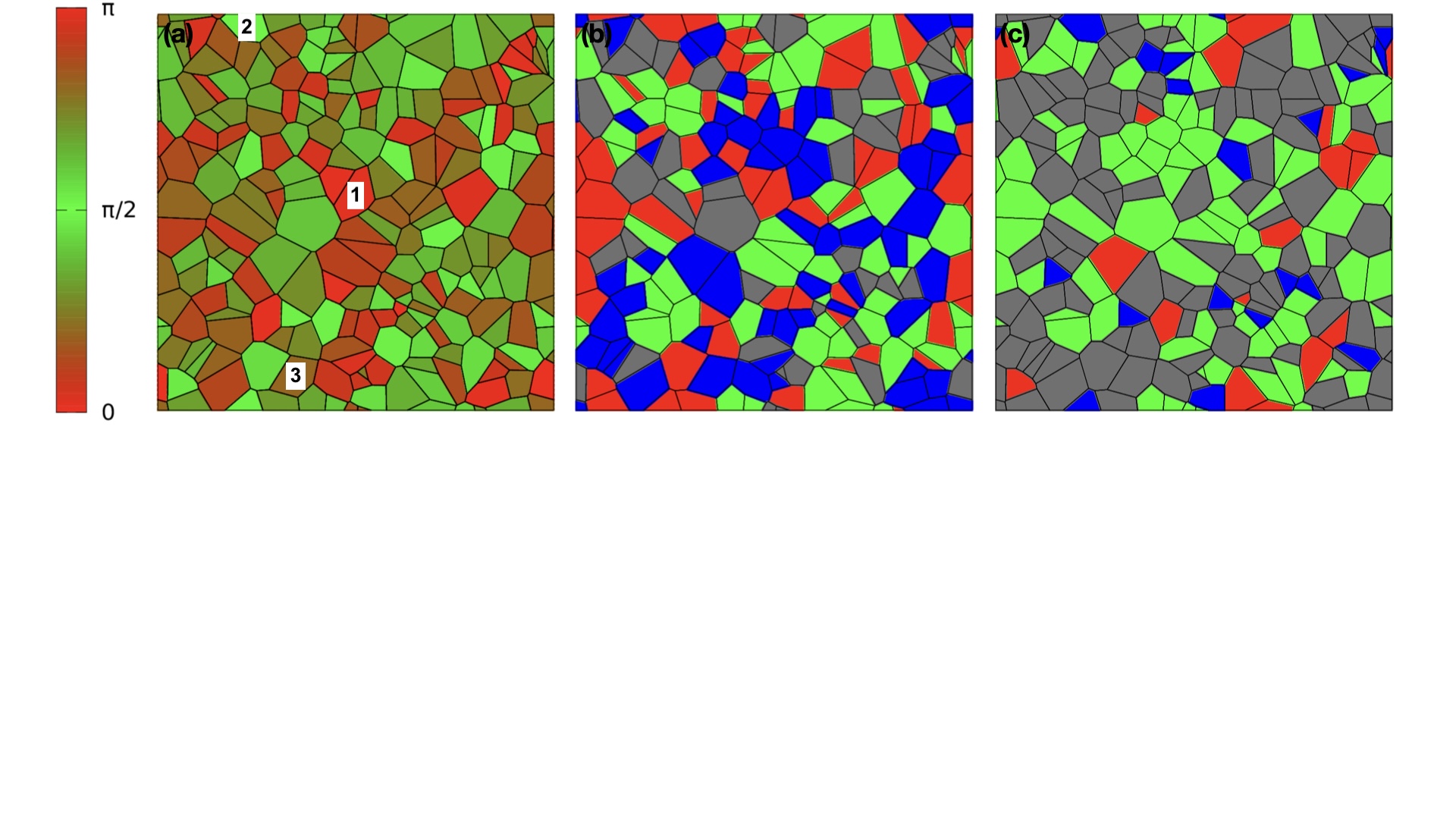}
\caption{Equilibrium variant distribution with random grain orientation.
\textbf{(a)} Grain orientation map corresponding to the microstructure. The grain rotation angle in the color bar is given in radian (modulo $\pi$ due to symmetry). The rotation axis is along the [001] direction.
The microstructure consists of 200 grains and tensile strain in horizontal direction is \textbf{(b)} $\langle\epsilon_{xx}\rangle/\epsilon_1 = 0$ and \textbf{(c)} $\langle\epsilon_{xx}\rangle/\epsilon_1 = 2.5$.
In all figures $\epsilon_0=\epsilon_1$, and the chosen Poisson ratio is $\nu=1/4$ (i.e. $\lambda=\mu$). The grains have a random orientation based on a uniform distribution.
}
\label{fig:rot}
\end{center}
\end{figure*}
The last example concerns a case where the grains are oriented randomly, i.e.~rotated  around the [001] axis, as depicted in the left panel of Fig.~\ref{fig:rot}.
Like in the previous publication \cite{Sandt2023a}, the response to a uniaxial strain $\langle \epsilon_{xx}\rangle$ is less pronounced than in the cases with uniform grain orientation above.
The center panel shows the equilibrium microstructure for vanishing mean strain, whereas the right panel is for a rather high tensile strain $\langle \epsilon_{xx}\rangle/\epsilon_1=2.5$.

For a clockwise rotation of a grain by an angle $\alpha$ around the [001] axis, the relevant components of the eigenstrain tensor become 
\begin{eqnarray}
\epsilon^{(0)'}_{xx} &=& \cos^2\alpha\, \epsilon_{xx}^{(0)} + 2\cos\alpha\sin\alpha\, \epsilon_{xy}^{(0)} + \sin^2\alpha\, \epsilon_{yy}^{(0)} \\
\epsilon^{(0)'}_{xy} &=& -\sin\alpha \cos\alpha\, \epsilon_{xx}^{(0)} + (\cos^2\alpha-\sin^2\alpha)\, \epsilon_{xy}^{(0)} \nonumber \\
&& + \sin\alpha\cos\alpha\, \epsilon_{yy}^{(0)} \\
\epsilon^{(0)'}_{yy} &=& \sin^2\alpha\, \epsilon_{xx}^{(0)} - 2\cos\alpha\sin\alpha\, \epsilon_{xy}^{(0)} + \cos^2\alpha\, \epsilon_{yy}^{(0)}.
\end{eqnarray}
Therefore, for the chosen example, a rotation by $\pi/4$ converts the tetragonal deformation to a shear transformation (and vice versa), apart from the eigenstrain contribution in $z$ direction, which only affects the energetic cost but not the strain distribution.
Therefore, we expect for a tensile load in [100] direction, that grains with orientation close to 0 or $\pi$ (the response is $\pi$-periodic), to have a tetragonal spin preferentially being in state $+1$;
for a rotation near $90^\circ$, the tetragonal contribution prefers the $s=-1$ state.
For both of these orientations (and nearby values), the shear contributions are indifferent, as they have only a vanishing or small contribution $\epsilon_{xx}^{(0)'}$, which is sensitive to the tensile strain.
As explained above, for rotations by approximately $\pi/4$ (or $3\pi/4$) the role of shear and tetragonal deformations is exchanged, but the original tetragonal transformation has an energy penalty due to the contribution in $z$ direction for the used plane strain setup.

For some selected grains in Fig.~\ref{fig:rot} the change of the variant states is discussed in the following. 
Grain 1 has an orientation close to $0$ (or $\pi$).
In the zero strain state, it is marked in red, hence the shear spin equals $-1$ and the tetragonal component is also $-1$.
Under stronger tensile load it turns grey, hence both spins flipped their sign, leading to a stronger alignment with the external field.
Grain 2 has an orientation close to $90^\circ$, is originally red (shear: -1, tetragonal -1) and becomes blue in the right panel.
In this case, the dominant effect is to have a negative tetragonal spin values, as discussed above.
Finally, for grain 3, which has a mixed orientation, we have effectively a transition from blue to grey, where the positive shear spin is relevant due to the orientation near $\pi/4$, whereas the flipping tetragonal spin component is less relevant.

\section{Conclusions}

From a materials science perspective, quantum quantum is only starting to influence the opportunities and methodologies of modeling approaches.
It can be expected, that in future general purpose quantum computing will have the potential to strongly accelerate materials science related simulations and to enable completely new possibilities. 
As of today, sufficiently large quantum computers are not yet available, and also suitable algorithms have not yet been developed.
Nevertheless, quantum annealing, as a specific type of adiabatic quantum computing, is commercially available with large numbers of qubits already today.
The type of problems which can be addressed is limited, as it is required to map the materials science related problem of interest to the minimization of an Ising Hamiltonian.

In the present work we have demonstrated, that the selection of multiple martensite variants can be mapped to an Ising problem for specific conditions and have demonstrated how the energy minimization can be performed efficiently via quantum annealing.
The superiority of QA compared to classical computing has been demonstrated in our previous publication \cite{Sandt2023a}.
The multiplicity of the domains together with the long-range, anisotropic and shape-dependant character of the elastic interactions make the determination of the ground state a particularly difficult problem, and therefore QA opens completely new possibilities. 
The central result of the present publication is the generalization to multi-variant systems and the analysis of the arising microstructures.
Also, we demonstrate how the proper coefficients of the Ising model can be obtained from general linear elastic calculations.

To illustrate the approach, we use a four variant per grain case, where each grain is represented by two Ising spins.
We use a linear combination of shear and tetragonal deformations to demonstrate the methodology.
As a result, the equilibrium microstructure depends on the applied external strains (or stresses), as well as on the strength of the distortions induced by the different variants.
The observed patterns can largely be explained by computations of the anisotropy of laminar arrangements of the different variants, showing that the mismatch of neighboring domains dominates the overall elastic energy.

For many applications in science, the formation of microstructures is critical and efficient simulation approaches to support experimental findings are highly desired.
Our developed quantum annealing approach presents an efficient new technological opportunity for the determination of equilibrium microstructures with long-range elastic interactions, where multiple martensite variants need to be considered.

\section{Acknowledgements}

This work was funded by the German Federal Ministry of Education and Research (BMBF) via the project ALANO (grant no.~13XP0396B) and the Helmholtz project ZeDaBase.
Open access was funded by the Deutsche Forschungsgemeinschaft (DFG, German Research Foundation) - 491111487.
The authors gratefully acknowledge the J\"ulich Supercomputing Centre (https://www.fz-juelich.de/ias/jsc) for funding this project by providing computing time on the D-Wave Advantage$^{\mathrm{TM}}$ System JUPSI through the J\"ulich UNified Infrastructure for Quantum computing (JUNIQ).


\begin{thebibliography}{10}

\bibitem{Karma1}
A.~Karma and W.~J. Rappel.
\newblock Phase-field method for computationally efficient modeling of
  solidification with arbitrary interface kinetics.
\newblock {\em Phys. Rev. E}, 53:R3017(R), 1996.

\bibitem{Karma2}
A.~Karma and W.~J. Rappel.
\newblock Quantitative phase-field modeling of dendritic growth in two and
  three dimension.
\newblock {\em Phys. Rev. E}, 57:4323, 1998.

\bibitem{Boussinot2014}
G.~Boussinot and E.~A. Brener.
\newblock Achieving realistic interface kinetics in phase-field models with a
  diffusional contrast.
\newblock {\em Phys. Rev. E}, 89:060402(R), 2014.

\bibitem{KaiWang2020}
K.~Wang, G.~Boussinot, C.~H\"uter, E.~A. Brener, and R.~Spatschek.
\newblock Modeling of dendritic growth using a quantitative nondiagonal phase
  field model.
\newblock {\em Phys. Rev. Mater.}, 4:033802, 2020.

\bibitem{Finel2018}
A.~Finel, Y.~{L}e Bouar, B.~Dabas, B.~Appolaire, Y.~Yamada, and T.~Mohri.
\newblock Sharp phase field method.
\newblock {\em Phys. Rev. Lett.}, 121:025501, 2018.

\bibitem{Sandt2023a}
R.~Sandt, Y.~Le Bouar, and R.~Spatschek.
\newblock Quantum annealing for microstructure equilibration with long-range
  elastic interactions.
\newblock {\em Sci. Rep.}, 13:6036, 2023.

\bibitem{Mao2019}
Binbin Yue, Fang Hong, Naohisa Hirao, Roman Vasin, Hans-Rudolf Wenk, Bin Chen,
  and Ho-Kwang Mao.
\newblock A simple variant selection in stress-driven martensitic
  transformation.
\newblock {\em Proc. Natl. Acad. Sci.}, 116(30):14905--14909, 2019.

\bibitem{Wechsler1953}
M.~Wechsler, D.~Lieberman, and T.~Read.
\newblock On the theory of the formation of martensite.
\newblock {\em Trans. AIME}, 197:1503--1515, 1953.

\bibitem{Bowles1954}
J.~S. Bowles and J.~K. Mackenzie.
\newblock The crystallography of martensite transformations {I}.
\newblock {\em Acta Metall.}, 2:129--137, 1954.

\bibitem{Finnila1994}
A.B. Finnila, M.A. Gomez, C.~Sebenik, C.~Stenson, and J.D. Doll.
\newblock Quantum annealing: a new method for minimizing multidimensional
  functions.
\newblock {\em Chem. Phys. Lett.}, 219:343--348, 1994.

\bibitem{Brooke1999}
J.~Brooke, D.~Bitko, T.F. Rosenbaum, and G.~Aeppli.
\newblock Quantum {A}nnealing of a {D}isordered {M}agnet.
\newblock {\em Science}, 284:779--781, 1999.

\bibitem{Kadowaki1998}
T.~Kadowaki and H.~Nishimori.
\newblock Quantum annealing in the transverse {I}sing model.
\newblock {\em Phys. Rev. E}, 58(5):5355--5363, 1998.

\bibitem{Morita2008}
S.~Morita and H.~Nishimori.
\newblock Mathematical foundation of quantum annealing.
\newblock {\em J. Math. Phys.}, 49:125210, 2008.

\bibitem{Rajak2023}
A.~Rajak, S.~Suzuki, A.~Dutta, and B.K. Chakrabarti.
\newblock Quantum annealing: an overview.
\newblock {\em Phil. Trans. R. Soc. A}, 381:20210417, 2023.

\bibitem{Warren2018}
R.H. Warren.
\newblock Mathematical methods for a quantum annealing computer.
\newblock {\em Adv. Appl. Math.}, 3(3):82--90, 2018.

\bibitem{Johnson2011}
M.W. Johnson, M.H. Amin, S.~Gildert, T.~Lanting, F.~Hamze, N.~Dickson,
  R.~Harris, A.J. Berkley, J.~Johansson, P.~Bunyk, E.M. Chapple, C.~Enderud,
  J.P. Hilton, K.~Karimi, E.~Ladizinsky, N.~Ladizinsky, T.~Oh, I.~Perminov,
  C.~Rich, M.C. Thom, E.~Tolkacheva, C.J.S. Truncik, S.~Uchaikin, J.~Wang,
  B.~Wilson, and G.~Rose.
\newblock Quantum annealing with manufactured spins.
\newblock {\em Nature}, 473:194--198, 2011.

\bibitem{Boixo2014}
S.~Boixo, T.F. R\o{}nnow, S.V. Isakov, Z.~Wang, D.~Wecker, D.A. Lidar, J.M.
  Martinis, and M.~Troyer.
\newblock Evidence for quantum annealing with more than one hundred qubits.
\newblock {\em Nat. Phys.}, 10:218--224, 2014.

\bibitem{Sandt2023b}
R.~Sandt and R.~Spatschek.
\newblock Efficient low temperature {M}onte {C}arlo sampling using quantum
  annealing.
\newblock {\em Sci. Rep.}, 13:6754, 2023.

\bibitem{Nelson2022a}
J.~Nelson, M.~Vuffray, A.Y. Lokhov, T.~Albash, and C.~Coffrin.
\newblock High-{Q}uality {T}hermal {G}ibbs {S}ampling with {Q}uantum
  {A}nnealing {H}ardware.
\newblock {\em Phys. Rev. Appl.}, 17:044046, 2022.

\bibitem{Nelson2022b}
M.~Vuffray, C.~Coffrin, Y.A. Kharkov, and A.Y. Lokhov.
\newblock Programmable {Q}uantum {A}nnealers as {N}oisy {G}ibbs {S}amplers.
\newblock {\em PRX Quantum}, 3:020317, 2022.

\bibitem{Mandra2017}
S.~Mandr\`{a}, Z.~Zhu, and H.G. Katzgraber.
\newblock Exponentially {B}iased {G}round-{S}tate {S}ampling of {Q}uantum
  {A}nnealing {M}achines with {T}ransverse-{F}ield {D}riving {H}amiltonians.
\newblock {\em Phys. Rev. Lett.}, 118:070502, 2017.

\bibitem{Koenz2019}
M.S. K\"onz, G.~Mazzola, A.J. Ochoa, H.G. Katzgraber, and M.~Troyer.
\newblock Uncertain fate of fair sampling in quantum annealing.
\newblock {\em Phys. Rev. A}, 100:030303, 2019.

\bibitem{Mukherjee2019}
S.~Mukherjee and B.~K. Chakrabarti.
\newblock On the question of ergodicity in quantum spin glass phase and its
  role in quantum annealing.
\newblock {\em J. Phys. Soc. Jpn.}, 88:061004, 2019.

\bibitem{Yamamoto2020}
M.~Yamamoto, M.~Ohzeki, and K.~Tanaka.
\newblock Fair {S}ampling by {S}imulated {A}nnealing on {Q}uantum {A}nnealer.
\newblock {\em J. Phys. Soc. Jpn.}, 89:025002, 2020.

\bibitem{Harris2018}
R.~Harris, Y.~Sato, A.J. Berkley, M.~Reis, F.~Altomare, M.H. Amin, K.~Boothby,
  P.~Bunyk, C.~Deng, C.~Enderud, S.~Huang, E.~Hoskinson, M.W. Johnson,
  E.~Ladizinsky, N.~Ladizinsky, T.~Lanting, R.~Li, T.~Medina, R.~Molavi,
  R.~Neufeld, T.~Oh, I.~Pavlov, I.~Perminov, G.~Poulin-Lamarre, C.~Rich,
  A.~Smirnov, L.~Swenson, N.~Tsai, M.~Volkmann, J.~Whittaker, and J.~Yao.
\newblock Phase transitions in a programmable quantum spin glass simulator.
\newblock {\em Science}, 361:162--165, 2018.

\bibitem{Kairys2020}
P.~Kairys, A.D. King, I.~Ozfidan, K.~Boothby, J.~Raymond, A.~Banerjee, and T.S.
  Humble.
\newblock Simulating the {S}hastry-{S}utherland {I}sing {M}odel {U}sing
  {Q}uantum {A}nnealing.
\newblock {\em PRX Quantum}, 1:020320, 2020.

\bibitem{Camino2023}
B.~Camino, J.~Buckeridge, P.~A. Warburton, V.~Kendon, and S.~M. Woodley.
\newblock Quantum computing and materials science: {A} practical guide to
  applying quantum annealing to the configurational analysis of materials.
\newblock {\em J. Appl. Phys.}, 133:221102, 2023.

\bibitem{Kitai2020}
K.~Kitai, J.~Guo, S.~Ju, S.~Tanaka, K.~Tsuda, J.~Shiomi, and R.~Tamura.
\newblock Designing metamaterials with quantum annealing and factorization
  machines.
\newblock {\em Phys. Rev. Res.}, 2:013319, 2020.

\bibitem{Juenger2021}
M.~J\"unger, E.~Lobe, P.~Mutzel, G.~Reinelt, F.~Rendl, G.~Rinaldi, and
  T.~Stollenwerk.
\newblock Quantum {A}nnealing versus {D}igital {C}omputing: {A}n {E}xperimental
  {C}omparison.
\newblock {\em ACM J. Exp. Algorithms}, 26(1):1.9, 2021.

\bibitem{Parekh2015}
O.~Parekh, J.~Wendt, L.~Shulenburger, A.~Landahl, J.~Moussa, and J.~Aidun.
\newblock Benchmarking {A}diabatic {Q}uantum {O}ptimization for {C}omplex
  {N}etwork {A}nalysis.
\newblock https://www.osti.gov/biblio/1459086, 2015.

\bibitem{Yan2022}
B.~Yan and N.A. Sinitsyn.
\newblock Analytical solution for nonadiabatic quantum annealing to arbitrary
  {I}sing spin {H}amiltonian.
\newblock {\em Nat. Commun.}, 13:2212, 2022.

\bibitem{King2023}
A.~D. King, J.~Raymond, T.~Lanting, R.~Harris, A.~Zucca, F.~Altomare, A.~J.
  Berkley, K.~Boothby, S.~Ejtemaee, C.~Enderud, E.~Hoskinson, S.~Huang,
  E.~Ladizinsky, A.~J.~R. MacDonald, G.~Marsden, R.~Molavi, T.~Oh,
  G.~Poulin-Lamarre, M.~Reis, C.~Rich, Y.~Sato, N.~Tsai, M.~Volkmann, J.~D.
  Whittaker, J.~Yao, A.~W. Sandvik, and M.~H. Amin.
\newblock Quantum critical dynamics in a 5,000-qubit programmable spin glass.
\newblock {\em Nature}, 617:61--66, 2023.

\bibitem{Sherrington2008}
D.~Sherrington.
\newblock A simple spin glass perspective on martensitic shape-memory alloys.
\newblock {\em J. Phys.: Condens. Matter}, 20:304213, 2008.

\bibitem{Kartha1991}
S.~Kartha, T.~Cast\'{a}n, J.A. Krumhansl, and J.P. Sethna.
\newblock Spin-{G}lass {N}ature of {T}weed {P}recursors in {M}artensitic
  {T}ransformations.
\newblock {\em Phys. Rev. Lett.}, 67:3630, 1991.

\bibitem{Sethna1992}
J.P. Sethna, S.~Kartha, T.~Cast\'{a}n, and J.A. Krumhansl.
\newblock Tweed in {M}artensites: {A} {P}otential {N}ew {S}pin {G}lass.
\newblock {\em Phys. Scr.}, 1992:214--219, 1992.

\bibitem{Vasseur2011}
R.~Vasseur and T.~Lookman.
\newblock Spin {M}odels for {F}erroelastics: {T}owards a {S}pin {G}lass
  {D}escription of {S}train {G}lass.
\newblock {\em Solid State Phenom.}, 172-174:1078--1083, 2011.

\bibitem{Polatidis2020}
E.~Polatidis, M.~Šmíd, I.~Kuběna, W.-N. Hsu, G.~Laplanche, and H.~Van
  Swygenhoven.
\newblock Deformation mechanisms in a superelastic {N}i{T}i alloy: {A}n in-situ
  high resolution digital image correlation study.
\newblock {\em Mater. Des.}, 191:108622, 2020.

\bibitem{Wang2022}
Z.~Wang, J.~Chen, R.~Kocich, I.~P.~Dolbnya S.~Tardif, L.~Kunčická, J.-S.
  Micha, K.~Liogas, O.~V. Magdysyuk, I.~Szurman, and A.~M. Korsunsky.
\newblock Grain {S}tructure {E}ngineering of {N}i{T}i {S}hape {M}emory {A}lloys
  by {I}ntensive {P}lastic {D}eformation.
\newblock {\em ACS Appl. Mater. Interfaces}, 14:31396--31410, 2022.

\bibitem{Manosa2016}
L.~Ma\~{n}osa and A.~Planes.
\newblock Mechanocaloric effects in shape memory alloys.
\newblock {\em Phil. Trans. R. Soc. A}, 374:20150310, 2016.

\bibitem{Otsuka2005}
K.~Otsuka and X.~Ren.
\newblock Physical metallurgy of {T}i–{N}i-based shape memory alloys.
\newblock {\em Prog. Mater. Sci.}, 50:511, 2005.

\bibitem{Ronnow2014}
T.F. R\o{}nnow, Z.~Wang, J.~Job, S.~Boixo, S.V. Isakov, D.~Wecker, J.M.
  Martinis, D.A. Lidar, and M.~Troyer.
\newblock Defining and detecting quantum speedup.
\newblock {\em Science}, 345:420--423, 2014.

\bibitem{Lucas2014}
A.~Lucas.
\newblock Ising formulations of many {NP} problems.
\newblock {\em Front. Phys.}, 2(5), 2014.

\bibitem{Lanting2014}
T.~Lanting, A.J Przybysz, A.Y. Smirnov, F.M. Spedalieri, M.H. Amin, A.J.
  Berkley, R.~Harris, F.~Altomare, S.~Boixo, P.~Bunyk, N.~Dickson, C.~Enderud,
  J.P. Hilton, E.~Hoskinson, M.W. Johnson, E.~Ladizinsky, N.~Ladizinsky,
  R.~Neufeld, T.~Oh, I.~Perminov, C.~Rich, M.C. Thom, E.~Tolkacheva,
  S.~Uchaikin, A.B. Wilson, and G.~Rose.
\newblock Entanglement in a {Q}uantum {A}nnealing {P}rocessor.
\newblock {\em Phys. Rev. X}, 4:021041, 2014.

\bibitem{Raymond2023}
J.~Raymond, R.~Stevanovic, W.~Bernoudy, K.~Boothby, C.C. McGeoch, A.J. Berkley,
  P.~Farr\'{e}, and A.D. King.
\newblock Hybrid quantum annealing for larger-than-{QPU} lattice-structured
  problems.
\newblock {\em ACM Trans. Quantum Comput.}, 2023.

\bibitem{Berwald2019}
J.J. Berwald.
\newblock The {M}athematics of {Q}uantum-{E}nabled {A}pplications on the
  {D}-{W}ave {Q}uantum {C}omputer.
\newblock {\em Not. Am. Math. Soc.}, 66(6):832--841, 2019.

\bibitem{Leap}
D-{W}ave {L}eap quantum cloud service.
\newblock {https://cloud.dwavesys.com}.
\newblock [Online; accessed 6-December-2023].

\bibitem{Khachaturyan2008}
A.G. Khachaturyan.
\newblock {\em Theory of {S}tructural {T}ransformations in {S}olids}.
\newblock Dover Publications, Inc., Mineola, New York, 2008.

\bibitem{Wang1998}
Y.~Wang, D.~Banerjee, C.~C. Su, and A.~G. Khachaturyan.
\newblock Field kinetic model and computer simulation of precipitation of
  {L}1$_2$ ordered intermetallics from f.c.c. solid solution.
\newblock {\em Acta Mater.}, 46:2983, 1998.

\bibitem{Lebbad2021}
H.~Lebbad, B.~Appolaire, Y.~Le~Bouar, and A.~Finel.
\newblock Insights into the selection mechanism of {W}idmanstätten growth by
  phase-field calculations.
\newblock {\em Acta Mater.}, 217:117148, 2021.

\bibitem{Yann1998}
Y.~Le~Bouar, A.~Loiseau, and A.G. Khachaturyan.
\newblock Origin of chessboard-like structures in decomposing alloys.
  {T}heoretical model and computer simulation.
\newblock {\em Acta Mater.}, 46:2777, 1998.

\end{thebibliography}

\end{document}